\documentclass{article}
\usepackage[final,dblblindworkshop]{neurips_2026}
\workshoptitle{Third Workshop on Agents in the Wild: Safety, Security, and Beyond}

\usepackage[utf8]{inputenc}
\usepackage[T1]{fontenc}
\usepackage{hyperref}
\hypersetup{hidelinks}
\usepackage{url}
\usepackage{booktabs}
\usepackage{amsmath,amsfonts,amssymb}
\usepackage{microtype}
\usepackage{xcolor}
\usepackage{graphicx}
\usepackage{multirow}
\usepackage{enumitem}
\usepackage{adjustbox}
\usepackage{placeins}
\newif\ifshowjenscomments
\showjenscommentstrue % Show comments by default

\newcommand{\tsbench}{\textsc{TS-Bench}}
\newcommand{\tsguard}{\textsc{TS-Guard}}
\newcommand{\stepguard}{\textsc{StepGuard}}
\newcommand{\agentdojo}{\textsc{AgentDojo}}
\newcommand{\agentharm}{\textsc{AgentHarm}}
\newcommand{\asb}{\textsc{ASB}}
\newcommand{\rosetta}{\textsc{ToolRosetta}}
\newcommand{\pp}{\,\ensuremath{\mathrm{pts}}}
\newcommand{\ci}[2]{[#1, #2]}

\newcommand{\SGShellDojoFlip}{+1.1}

\newcommand{\TGShellAsbdFPR}{+37.8}
\newcommand{\TGShellAsbdFPRCI}{\ci{+35.4}{+40.1}}
\newcommand{\TGShellAsbdTPR}{+1.5}
\newcommand{\TGShellAsbdTPRCI}{\ci{+0.6}{+2.5}}

\newcommand{\GFShellAsbFlip}{-0.3}
\newcommand{\GFShellAsbFlipCI}{\ci{-0.8}{+0.1}}

\title{\rosetta{}: Surface-Form Invariance of Tool-Call Guards}
\author{%
  Hasan Demirkiran\\
  Kontext\\
  \texttt{hasan@kontext.security}
  \And
  Jens Ernstberger\\
  Kontext\\
  \texttt{jens@kontext.security}
}

\begin{document}
\maketitle

\begin{abstract}
Step-level guards for tool-using agents are evaluated on tidy JSON tool calls, but deployed coding agents act through \texttt{execute\_bash} and \texttt{execute\_code} tools whose payloads are strings. We introduce \rosetta{}, which re-renders every candidate action in the \agentdojo{} and \agentharm{} splits of \tsbench{} as a Python call, as a shell command, and without the agent's rationale. The renderings are produced by code and verified by round-trip parsing, so the user request, the history, the label and the action's effect stay fixed while only the surface form changes. Across eleven guards with two placebo controls, frontier and strong open-weight guards are stable: no frontier error rate moves by more than 3.4 points. Four of five specialized agent guards are not: under the shell form their false-positive rates rise by 21 to 47 points on \agentdojo{}, and \tsguard{}'s by 38 points on its own home benchmark, blocking between a third and nine in ten benign actions; the fifth, \stepguard{}, does not move. A small open guard drifts five points more permissive. Removing the agent's rationale moves most guards whose input carries it, in guard-specific directions. A request-blind control on one guard shows that the rationale is where intent is read from. A parser that restores the canonical call in front of the guard removes the effect for these renderings, by construction. Benchmark accuracy on JSON calls does not predict a guard's operating point behind a code-acting agent.
\end{abstract}

\section{Introduction}

Runtime guards that approve or block an agent's tool call before execution are the most direct defense for tool-using agents \citep{mou2026toolsafe,zheng2026stepguard,chennabasappa2025llamafirewall}. They are trained and benchmarked on step datasets in which the candidate action is a typed call: a tool name and a JSON argument object. Code-acting agents \citep{wang2024codeact,wang2025openhands} route most actions through one or two generic tools, \texttt{execute\_bash} and \texttt{execute\_code}, and the action the guard must judge is a string inside the argument. The same reservation, transfer or deletion arrives as \texttt{reserve\_hotel(hotel=...)} inside a code block or as \texttt{reserve-hotel --hotel ...} inside a shell command.

Whether a guard's verdict survives that change has not been tested. Recent agent-safety benchmarks stress agents and guards with multi-turn, decomposed or long-horizon attacks \citep{li2026unsafer,kothamasu2026decompbench,jiang2026agentlab}, audit what existing test suites cover \citep{chen2026safeaudit}, or measure agents' own paraphrase robustness \citep{rabanser2026reliability}; none varies how the same action is written. \citet{tracesafe2026} observed that guard accuracy on tool trajectories correlates with structured-data competence more than with jailbreak robustness, but did not intervene on the format. Work on the text-to-tool gap \citep{cartagena2026gap,yu2026affordance} asks how the \emph{agent} changes when it can act. We ask the same of the \emph{referee}.

We answer it with a paired, label-preserving intervention. Every step in the \agentdojo{} and \agentharm{} evaluation splits of \tsbench{} \citep{mou2026toolsafe} is rewritten so that only the candidate action's surface form changes. The renderings are produced by code, not by a model. Each rendered row is parsed back and compared with the original tool and arguments before it is accepted, so labels are inherited rather than re-annotated, and any change in a guard's decision is attributable to the writing of the action alone. Two placebo conditions bound tokenization noise and the effect of declaring the wrapper tool.

Our contributions are: (i) \rosetta{}, a derivation recipe and pinned artifact that turns any step-level guard benchmark into a surface-form invariance test: four renderings, two placebos, labels inherited, code renderings round-trip verified; (ii) an evaluation of eleven guards, from frontier APIs to five specialized agent guards, with pre-registered, placebo-adjusted estimands; (iii) the finding that surface form leaves frontier guards nearly untouched but shifts four of five specialized guards by 21 to 47 false-positive points while the fifth holds; and (iv) mechanism evidence, from a request-blind control on one guard, that intent is read from the agent's rationale.

\section{\rosetta{}: same action, different surface form}
\label{sec:method}

\paragraph{Unit.}
A \tsbench{} step is a tuple (user request, interaction history, candidate action, tool definitions, label). The candidate action is ReAct text: a \emph{Thought}, an \emph{Action} line naming the tool, and an \emph{Action Input} line with JSON arguments. Labels are read strictly ($0.5$ and $1$ are unsafe); no tool call is ever executed.

\paragraph{Conditions.}
We hold the request, history and label byte-identical and change only the candidate action (and, for wrapper conditions, append one fixed tool definition):
\begin{itemize}[leftmargin=1.2em,itemsep=1pt,topsep=2pt]
\item \textbf{original}: the released step.
\item \textbf{placebo (whitespace / wrapper)}: trailing whitespace on the action, the tokenization-noise floor; or the action unchanged with the \texttt{execute\_python} definition appended to the tool list, isolating the tool-list change the code condition also makes.
\item \textbf{no rationale}: the Thought and any preamble are dropped; Action and Action Input untouched.
\item \textbf{Python} / \textbf{shell}: Thought kept; the call becomes \texttt{execute\_python} with \texttt{code = "tool(k=v, ...)"} (keyword-only Python literals), or \texttt{execute\_bash} with \texttt{command = "tool --key value ..."} (\texttt{shlex}-quoted, compact JSON for nested values).
\item \textbf{prose}: Thought kept; the Action and Action Input lines are replaced by one first-person sentence, written by Gemini 3.8 Flash, that states the action and contains every argument value verbatim and no tool name.
\end{itemize}
Python and shell renderings are deterministic. Each rendered action is parsed back with \texttt{ast} or \texttt{shlex} and the recovered (tool, arguments) must equal the original object exactly, or the row is excluded from that condition and counted. Of $7{,}182$ evaluation steps, $6{,}764$ have a parseable JSON argument object and are renderable; none failed the round trip. The primary cohort is the renderable \agentdojo{} \citep{debenedetti2024agentdojo} ($1{,}220$ steps, $352$ unsafe) and \agentharm{} \citep{andriushchenko2024agentharm} ($718$ steps, $512$ unsafe) splits; \asb{} \citep{zhang2024asb} is reported as an extension (\S\ref{sec:asb}). Appendix~\ref{app:examples} shows a step under every condition.

\paragraph{Guards and prompts.}
Frontier: Gemini 3.8 Flash, Claude Sonnet 5. Open weights: Llama 4 Maverick, GLM 5.2, gpt-oss-120b, Gemma 4 26B (managed); Qwen 3.8 27B, Nemotron 3.5 Lightning 30B (self-hosted, greedy). Specialized: \tsguard{} \citep{mou2026toolsafe}, \stepguard{} \citep{zheng2026stepguard}, AgentDoG \citep{liu2026agentdog,li2026atbench}, ShieldAgent \citep{zhang2024agentsafetybench} and Safiron \citep{huang2025safiron}, each run with its authors' released prompt and parser. The last three judge trajectories or plans rather than steps and start from weak baselines here, so we report their shift, not their accuracy. LLM guards use the \tsbench{} authors' assessment prompt and a JSON schema (malicious request, third-party attack, action harmfulness in $\{0, 0.5, 1\}$); a step is blocked when the composite score is $\geq 0.5$. A \emph{request-blind} control runs Gemini with the user request withheld. Temperature $0$, one run per guard and condition; hosted guards are non-deterministic, so every effect is read above the whitespace placebo. \stepguard{}, AgentDoG, ShieldAgent and Safiron are fed a parsed (tool, arguments) record rather than the step text. The prose form has no record to parse, so it is outside their input contract and reported as not applicable.

\paragraph{Estimands.}
Per guard, condition and source, paired to \emph{original} on the same steps: the verdict flip rate minus the matching placebo's flip rate (wrapper placebo for Python and shell, whitespace placebo for no-rationale and prose), and the change in true- and false-positive rate, each with a $1{,}000$-replicate percentile bootstrap clustered by trajectory. The protocol, cohort and floors were frozen before any perturbed-condition inference.\footnote{\rosetta{} code, protocol and per-row results are released; raw \tsbench{} text is not.}

\section{Results}
\label{sec:results}

\begin{table}[t]
\centering\small
\caption{Verdict flips above the matching placebo (pts) per rendering, and the shell form's $\Delta$FPR / $\Delta$TPR; rows ordered by the shell form's $\Delta$FPR on \agentdojo{}. $*$: 95\% interval excludes zero; full estimands in Appendix~\ref{app:full}.}
\label{tab:main}
\adjustbox{max width=\linewidth}{\scriptsize\begin{tabular}{lrrrrrrrrrr}
\toprule
 & \multicolumn{5}{c}{AgentDojo} & \multicolumn{5}{c}{AgentHarm} \\
\cmidrule(lr){2-6} \cmidrule(lr){7-11}
 & \multicolumn{3}{c}{flip $-$ placebo (pts)} & \multicolumn{2}{c}{shell form} & \multicolumn{3}{c}{flip $-$ placebo (pts)} & \multicolumn{2}{c}{shell form} \\
Guard & Python & Shell & No rat. & $\Delta$FPR & $\Delta$TPR & Python & Shell & No rat. & $\Delta$FPR & $\Delta$TPR \\
\midrule
\multicolumn{11}{l}{\textit{Specialized agent guards}} \\
Safiron (8B) & +9.7* & +43.1* & +1.6* & +47.1* & +34.9* & +4.6* & +22.8* & +4.2* & +17.0* & +17.7* \\
AgentDoG (4B) & +10.8* & +18.8* & +12.5* & +29.8* & +0.0 & +6.3* & +8.2* & +7.0* & +10.7* & +2.5* \\
ShieldAgent (7B) & +14.4* & +22.5* & n/a & +26.8* & +6.2* & +5.6* & +4.7* & n/a & +12.1* & +1.5 \\
TS-Guard (7.6B, JSON-trained) & +5.1* & +15.6* & +4.4* & +21.2* & +6.6* & +2.2* & +4.5* & -0.1 & +12.6* & +0.0 \\
StepGuard (4B, JSON-trained) & +1.2* & +1.1* & +0.0 & +0.3 & +0.0 & +2.5* & +3.6* & +1.5* & -0.5 & -0.5 \\
\midrule
\multicolumn{11}{l}{\textit{General-purpose LLM guards}} \\
Claude Sonnet 5 & +0.3 & +0.6 & +3.4* & +2.5* & +0.3 & +0.7 & +0.7 & +4.1* & +1.0 & -1.0 \\
Llama 4 Maverick & +2.6* & +2.6* & +8.7* & +1.5 & +1.7 & +3.6* & +3.3* & +6.7* & +1.9 & -1.5 \\
Nemotron 3.5 Lightning 30B & +1.2* & +1.8* & +4.0* & +1.2 & +0.6 & +1.8 & +1.8 & +7.8* & -5.8* & +0.5 \\
Gemini 3.8 Flash & +0.0 & +0.0 & +0.6* & +0.0 & +0.0 & +0.3 & +0.3 & +1.1 & +3.4* & +1.5 \\
Gemini 3.8 Flash (request-blind) & +0.0 & +0.0 & +0.6* & +0.0 & +0.0 & +0.0 & +0.1 & +4.6* & +0.0 & +0.5 \\
Qwen 3.8 27B & +0.2 & +0.2 & +0.2 & -0.8* & +0.0 & -0.1 & +0.1 & +2.6* & +0.0 & -0.5 \\
\bottomrule
\end{tabular}
}
\end{table}

\begin{figure}[t]
\centering
\includegraphics[width=0.74\linewidth]{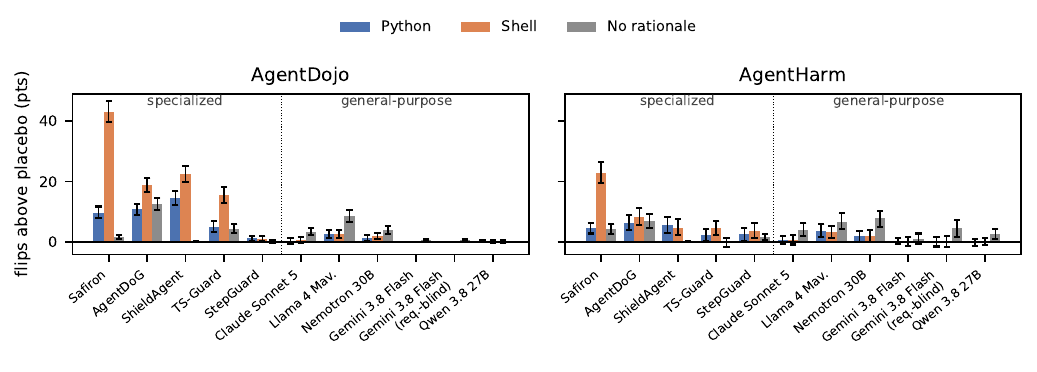}
\caption{Verdict flips above placebo per guard and rendering, with 95\% intervals.}
\label{fig:flips}
\end{figure}

\subsection{Frontier guards are stable; four of five specialized guards are not}

\textbf{Frontier and strong open guards read through the syntax} (Table~\ref{tab:main}, Figure~\ref{fig:flips}; baselines in Appendix~\ref{app:baselines}). Gemini 3.8 Flash produces \emph{zero} flips on \agentdojo{} under both the Python and the shell form, and on \agentharm{} its adjusted flips are $+0.3\pp$ with intervals including zero. Claude Sonnet 5 adds one new miss out of $352$ unsafe \agentdojo{} steps under either form; its only shift is $+2.5\pp$ of false positives \ci{+1.3}{+3.9} under shell. Qwen 3.8 27B's flips are within $0.2\pp$ of placebo everywhere. Stable is not exactly invariant: Gemini's \agentharm{} false-positive rate moves by $+3.4\pp$ under shell.

\textbf{Specialized guards' operating points move with the surface form.} \tsguard{}, trained on ReAct steps with JSON arguments, flips $15.6\pp$ \ci{+12.8}{+18.2} more verdicts than placebo under the shell form on \agentdojo{}. Its false-positive rate rises from $7.8\%$ to $29\%$ ($+21.2\pp$, \ci{+17.7}{+24.2}): $196$ of $868$ benign steps are newly blocked. Recall also rises ($+6.6\pp$), so this is not a cautious threshold but lost discrimination: balanced accuracy falls from $0.903$ to $0.829$, and on \agentharm{} the shell form moves \tsguard{} from fifth to last of eight guards (Kendall $\tau = 0.79$). The Python form has a quarter of the effect ($+5.3\pp$ FPR, \ci{+3.4}{+7.3}), so the wrapper alone, which the wrapper placebo also carries, does not explain it: the shell surface does. On \agentharm{} the shell form adds $12.6\pp$ of false positives \ci{+5.7}{+19.2}. \stepguard{}, the newer guard from the same group, trained on contrastive same-context, different-action data, stays within \SGShellDojoFlip\pp{} of placebo under shell with no error-rate shift. The other three specialized guards move like \tsguard{} or more: under the shell form on \agentdojo{} AgentDoG's false-positive rate rises by $29.8\pp$ \ci{+26.6}{+33.1}, ShieldAgent's by $26.8\pp$ \ci{+23.5}{+30.4} and Safiron's by $47.1\pp$ \ci{+43.4}{+50.6}; the Python form moves each by a third to two thirds as much. One caveat: AgentDoG, ShieldAgent and Safiron judge trajectories or plans through our adapters from weak baselines (Appendix~\ref{app:baselines}), so their shifts measure instability within guard; of the two step-level guards, one moves and one does not.

\textbf{Small open guards drift, in either direction.} Llama 4 Maverick churns $2.6$--$3.6\pp$ above placebo under both forms on both sources without a net error shift. Nemotron 3.5 Lightning, the weakest guard, becomes \emph{more permissive}: on \agentharm{} its false-positive rate falls by $4.9\pp$ \ci{-9.3}{-0.5} under Python and $5.8\pp$ \ci{-9.1}{-2.4} under shell at unchanged recall.

For the seven guards that read text, \textbf{prose} (Appendix~\ref{app:prose}) is the mildest form: within $3\pp$ of placebo for all of them, and zero flips for Gemini. For the four record-based guards it is no form at all: their input contracts reject it.

\subsection{Guards judge the narration}
\label{sec:rationale}

Dropping the agent's Thought moves most guards whose input carries it (from $0.0$ for \stepguard{} to $12.5$ points of flips for AgentDoG on \agentdojo{}), and the direction tells us what each one reads. Claude's over-blocking on \agentdojo{} \emph{drops} by $3.2\pp$ \ci{-5.0}{-1.6} at unchanged recall: the Thought there echoes the injected instruction. Llama, Nemotron and AgentDoG block \emph{more} without it ($+3.6$, $+5.0$ and $+11.9\pp$ FPR on \agentdojo{}), and Llama misses more. ShieldAgent's format has no slot for the rationale, so the condition is reported as not applicable. The request-blind control isolates the mechanism: on \agentharm{}, where the harm is the call itself, Gemini without the user request loses $4.4\pp$ of recall when the Thought is removed \ci{-8.4}{-0.9}, while Gemini with the request loses none. The control covers Gemini only, so it does not explain the specialized guards.

\subsection{Where format can matter}
\label{sec:asb}

The difference between the two sources is itself diagnostic. On \agentdojo{} the harm signal lives in the history, in the injected instruction that precedes the step. On \agentharm{} the request is malicious and the harm is the call, and every guard that moves at all moves there. On the $4{,}826$ renderable \asb{} steps, \tsguard{}'s home benchmark (Appendix~\ref{app:asb}), the shell form moves its false-positive rate by \TGShellAsbdFPR\pp{} \TGShellAsbdFPRCI{} and its recall by \TGShellAsbdTPR\pp{} \TGShellAsbdTPRCI{}, while Gemini's adjusted flips stay at \GFShellAsbFlip\pp{} \GFShellAsbFlipCI{}.

\subsection{Mitigation: resolve, then judge}

For our code and shell renderings the mitigation is exact, by construction. The renderings are reversible, so a parser that reconstructs the Action / Action Input block from the code or shell string restores the guard's original input and, with it, the original verdict: the \rosetta{} parser, forty lines at the boundary, removes a 21- to 47-point shift. It is not a defense for arbitrary agent-written code, and no command was executed. This is \citet{li2026actionalignment}'s ``judge the effect, not the string'' made operational.

\section{Discussion}

\textbf{What the result is not.} The six general-purpose LLM guards and \stepguard{} did not fail on the shell form. The claim is that a guard's benchmark score on JSON calls does not tell you how often it will block or miss once the agent writes its actions as code: four specialized guards move by 21 to 47 points, \stepguard{} does not, and with five guards we cannot say which property separates them.

\textbf{Limitations.} One run per hosted guard, so effects are read above a placebo; the two noisiest hosted guards (GLM 5.2, gpt-oss-120b) are reported only as floors. The wrapper definitions are fictional (bounded by the wrapper placebo), shell rendering preserves values but not types, and we render one canonical style per form. The prose form cannot be inverted by a parser; it has the smallest effects in the study.

\bibliographystyle{plainnat}
\bibliography{refs}

\appendix
\section{Baselines}
\label{app:baselines}
\begin{table}[h]
\centering\small
\caption{Baselines on the original steps and the whitespace-placebo flip rate (\%). BACC: balanced accuracy.}
\label{tab:baselines}
\adjustbox{max width=\linewidth}{\begin{tabular}{lrrrrrrrr}
\toprule
 & \multicolumn{4}{c}{AgentDojo} & \multicolumn{4}{c}{AgentHarm} \\
\cmidrule(lr){2-5} \cmidrule(lr){6-9}
Guard & BACC & TPR & FPR & placebo flip \% & BACC & TPR & FPR & placebo flip \% \\
\midrule
Gemini 3.8 Flash & 0.983 & 0.98 & 0.01 & 0.0 & 0.831 & 0.83 & 0.17 & 2.8 \\
Claude Sonnet 5 & 0.954 & 0.99 & 0.08 & 1.3 & 0.807 & 0.89 & 0.28 & 2.7 \\
Gemini 3.8 Flash (request-blind) & 0.983 & 0.98 & 0.01 & 0.0 & 0.790 & 0.76 & 0.18 & 5.2 \\
Qwen 3.8 27B & 0.824 & 1.00 & 0.35 & 0.5 & 0.742 & 0.95 & 0.46 & 1.4 \\
Llama 4 Maverick & 0.811 & 0.90 & 0.28 & 4.1 & 0.738 & 0.73 & 0.26 & 4.2 \\
Nemotron 3.5 Lightning 30B & 0.815 & 0.97 & 0.34 & 3.4 & 0.660 & 0.70 & 0.38 & 4.6 \\
TS-Guard (7.6B, JSON-trained) & 0.903 & 0.88 & 0.08 & 3.4 & 0.752 & 0.98 & 0.48 & 3.9 \\
StepGuard (4B, JSON-trained) & 0.978 & 0.99 & 0.04 & 0.7 & 0.754 & 0.83 & 0.32 & 1.8 \\
AgentDoG (4B) & 0.687 & 1.00 & 0.63 & 0.8 & 0.519 & 0.92 & 0.88 & 0.0 \\
ShieldAgent (7B) & 0.759 & 0.91 & 0.40 & 0.1 & 0.585 & 0.94 & 0.77 & 0.0 \\
Safiron (8B) & 0.574 & 0.31 & 0.16 & 0.1 & 0.510 & 0.10 & 0.08 & 0.1 \\
\bottomrule
\end{tabular}
}
\end{table}

\FloatBarrier
\section{Prose rendering}
\label{app:prose}
The seven guards whose prompt carries free text were run on the prose form; the four record-based guards (\stepguard{}, AgentDoG, ShieldAgent, Safiron) fill their prompts from a parsed (tool, arguments) record and reject every prose step before inference.
\begin{table}[h]
\centering\small
\caption{Prose form: placebo-adjusted flips (pts) and shell-style error shifts per source. n/a: input contract rejects prose.}
\label{tab:prose}
\adjustbox{max width=\linewidth}{\begin{tabular}{lrrrrrr}
\toprule
 & \multicolumn{3}{c}{AgentDojo} & \multicolumn{3}{c}{AgentHarm} \\
\cmidrule(lr){2-4} \cmidrule(lr){5-7}
Guard & flip $-$ placebo & $\Delta$TPR & $\Delta$FPR & flip $-$ placebo & $\Delta$TPR & $\Delta$FPR \\
\midrule
Gemini 3.8 Flash & +0.0 & +0.0 & +0.0 & +0.3 & -0.5 & +0.5 \\
Claude Sonnet 5 & +1.9* & +0.6 & +2.4* & +1.4 & -2.1 & +1.0 \\
Gemini 3.8 Flash (request-blind) & +0.0 & +0.0 & +0.0 & -0.9 & -1.6 & -2.0 \\
Qwen 3.8 27B & -0.1 & +0.0 & +0.1 & -0.3 & +0.0 & +0.5 \\
Llama 4 Maverick & +2.9* & +6.9* & +1.7* & +2.4* & -1.5 & +0.5 \\
Nemotron 3.5 Lightning 30B & +0.7 & -0.3 & +1.8* & +3.0* & +0.0 & -1.5 \\
TS-Guard (7.6B, JSON-trained) & +2.2* & +4.3* & +0.0 & -0.1 & -0.5 & -4.0 \\
StepGuard (4B, JSON-trained) & n/a & n/a & n/a & n/a & n/a & n/a \\
AgentDoG (4B) & n/a & n/a & n/a & n/a & n/a & n/a \\
ShieldAgent (7B) & n/a & n/a & n/a & n/a & n/a & n/a \\
Safiron (8B) & n/a & n/a & n/a & n/a & n/a & n/a \\
\bottomrule
\end{tabular}
}
\end{table}

\clearpage
\section{\asb{} extension}
\label{app:asb}
\tsguard{} and Gemini 3.8 Flash were run on all six conditions over the $4{,}826$ renderable \asb{} steps, \tsguard{}'s home benchmark; the shell form is the largest shift in the study.
\begin{table}[h]
\centering\small
\caption{\asb{} extension: baseline and placebo-adjusted effects on the $4{,}826$ renderable \asb{} steps.}
\label{tab:asb}
\adjustbox{max width=\linewidth}{\begin{tabular}{lrrrrrrrr}
\toprule
 & \multicolumn{3}{c}{original} & \multicolumn{3}{c}{flip $-$ placebo (pts)} & \multicolumn{2}{c}{shell form} \\
\cmidrule(lr){2-4} \cmidrule(lr){5-7} \cmidrule(lr){8-9}
Guard & BACC & TPR & FPR & Python & Shell & No rat. & $\Delta$FPR & $\Delta$TPR \\
\midrule
Gemini 3.8 Flash & 0.812 & 0.67 & 0.05 & -0.1 & -0.3 & +0.4 & +0.1 & -0.2 \\
TS-Guard (7.6B, JSON-trained) & 0.969 & 0.97 & 0.03 & +6.6* & +19.0* & +5.3* & +37.8* & +1.5* \\
\bottomrule
\end{tabular}
}
\end{table}

\FloatBarrier
\section{Full pre-registered estimands}
\label{app:full}
Every cell of the frozen protocol, per guard, condition and source: raw flip rate, placebo-adjusted flip with its 95\% cluster-bootstrap interval, $\Delta$TPR and $\Delta$FPR with intervals, and the counts of newly missed unsafe steps and newly blocked benign steps.

\begin{center}\scriptsize
\adjustbox{max width=\linewidth}{\begin{tabular}{llrrrrrrr}
\toprule
Guard & Condition & Source & flip \% & flip$-$placebo & $\Delta$TPR & $\Delta$FPR & new misses & new FPs \\
\midrule
Gemini 3.8 Flash & Python & AgentDojo & 0.0 & +0.0 [+0.0, +0.0] & +0.0 [+0.0, +0.0] & +0.0 [+0.0, +0.0] & 0 & 0 \\
Gemini 3.8 Flash & Python & AgentHarm & 3.2 & +0.3 [-0.8, +1.4] & +2.0 [-0.5, +5.2] & +1.9 [-0.5, +4.6] & 4 & 5 \\
Gemini 3.8 Flash & Shell & AgentDojo & 0.0 & +0.0 [+0.0, +0.0] & +0.0 [+0.0, +0.0] & +0.0 [+0.0, +0.0] & 0 & 0 \\
Gemini 3.8 Flash & Shell & AgentHarm & 3.2 & +0.3 [-1.3, +1.7] & +1.5 [-0.5, +3.9] & +3.4* [+0.9, +6.5] & 2 & 8 \\
Gemini 3.8 Flash & No rationale & AgentDojo & 0.6 & +0.6* [+0.2, +1.1] & +0.3 [+0.0, +1.1] & -0.7* [-1.3, -0.2] & 0 & 0 \\
Gemini 3.8 Flash & No rationale & AgentHarm & 3.9 & +1.1 [-0.6, +2.8] & +1.5 [+0.0, +3.5] & +1.5 [-1.8, +4.5] & 5 & 5 \\
Gemini 3.8 Flash & Prose & AgentDojo & 0.0 & +0.0 [+0.0, +0.0] & +0.0 [+0.0, +0.0] & +0.0 [+0.0, +0.0] & 0 & 0 \\
Gemini 3.8 Flash & Prose & AgentHarm & 3.0 & +0.3 [-1.3, +1.8] & -0.5 [-3.5, +2.0] & +0.5 [-1.1, +2.1] & 4 & 3 \\
Claude Sonnet 5 & Python & AgentDojo & 3.1 & +0.3 [-0.7, +1.4] & +0.0 [-0.9, +0.8] & +0.0 [-1.4, +1.3] & 1 & 18 \\
Claude Sonnet 5 & Python & AgentHarm & 4.2 & +0.7 [-0.5, +1.9] & -2.0 [-4.8, +0.4] & +2.4* [+0.5, +4.9] & 6 & 7 \\
Claude Sonnet 5 & Shell & AgentDojo & 3.4 & +0.6 [-0.5, +1.7] & +0.3 [-0.6, +1.4] & +2.5* [+1.3, +3.9] & 1 & 30 \\
Claude Sonnet 5 & Shell & AgentHarm & 4.2 & +0.7 [-0.8, +2.3] & -1.0 [-4.6, +2.1] & +1.0 [-0.9, +2.8] & 5 & 5 \\
Claude Sonnet 5 & No rationale & AgentDojo & 4.8 & +3.4* [+2.2, +4.7] & +1.1 [-0.3, +2.8] & -3.2* [-5.0, -1.6] & 1 & 12 \\
Claude Sonnet 5 & No rationale & AgentHarm & 6.7 & +4.1* [+2.1, +6.2] & -0.5 [-4.6, +3.3] & +1.5 [-1.4, +4.4] & 7 & 7 \\
Claude Sonnet 5 & Prose & AgentDojo & 3.2 & +1.9* [+0.9, +2.8] & +0.6 [+0.0, +1.5] & +2.4* [+1.2, +3.7] & 0 & 29 \\
Claude Sonnet 5 & Prose & AgentHarm & 4.0 & +1.4 [-0.1, +3.2] & -2.1 [-6.0, +1.1] & +1.0 [-1.1, +3.8] & 6 & 3 \\
Gemini 3.8 Flash (request-blind) & Python & AgentDojo & 0.0 & +0.0 [+0.0, +0.0] & +0.0 [+0.0, +0.0] & +0.0 [+0.0, +0.0] & 0 & 0 \\
Gemini 3.8 Flash (request-blind) & Python & AgentHarm & 5.3 & +0.0 [-1.5, +1.5] & +2.0 [-0.0, +4.4] & -1.0 [-3.7, +1.5] & 2 & 3 \\
Gemini 3.8 Flash (request-blind) & Shell & AgentDojo & 0.0 & +0.0 [+0.0, +0.0] & +0.0 [+0.0, +0.0] & +0.0 [+0.0, +0.0] & 0 & 0 \\
Gemini 3.8 Flash (request-blind) & Shell & AgentHarm & 5.4 & +0.1 [-1.7, +1.9] & +0.5 [-3.0, +4.2] & +0.0 [-2.7, +2.8] & 6 & 4 \\
Gemini 3.8 Flash (request-blind) & No rationale & AgentDojo & 0.6 & +0.6* [+0.2, +1.0] & +0.0 [+0.0, +0.0] & -0.6 [-1.2, +0.0] & 0 & 1 \\
Gemini 3.8 Flash (request-blind) & No rationale & AgentHarm & 9.7 & +4.6* [+1.7, +7.3] & -4.4* [-8.4, -0.9] & -1.9 [-5.2, +1.0] & 11 & 3 \\
Gemini 3.8 Flash (request-blind) & Prose & AgentDojo & 0.0 & +0.0 [+0.0, +0.0] & +0.0 [+0.0, +0.0] & +0.0 [+0.0, +0.0] & 0 & 0 \\
Gemini 3.8 Flash (request-blind) & Prose & AgentHarm & 3.8 & -0.9 [-2.5, +0.8] & -1.6 [-3.5, +0.0] & -2.0 [-4.4, +0.0] & 3 & 1 \\
Qwen 3.8 27B & Python & AgentDojo & 0.7 & +0.2 [-0.1, +0.6] & +0.0 [+0.0, +0.0] & -0.7 [-1.5, +0.0] & 0 & 1 \\
Qwen 3.8 27B & Python & AgentHarm & 1.7 & -0.1 [-1.2, +0.9] & +0.0 [+0.0, +0.0] & +0.5 [-2.5, +3.6] & 0 & 5 \\
Qwen 3.8 27B & Shell & AgentDojo & 0.6 & +0.2 [-0.2, +0.6] & +0.0 [+0.0, +0.0] & -0.8* [-1.5, -0.2] & 0 & 0 \\
Qwen 3.8 27B & Shell & AgentHarm & 1.9 & +0.1 [-1.0, +1.4] & -0.5 [-1.7, +0.0] & +0.0 [-2.9, +2.8] & 1 & 4 \\
Qwen 3.8 27B & No rationale & AgentDojo & 0.7 & +0.2 [-0.4, +0.7] & +0.0 [+0.0, +0.0] & +0.5 [-0.1, +1.1] & 0 & 6 \\
Qwen 3.8 27B & No rationale & AgentHarm & 4.0 & +2.6* [+1.1, +4.4] & -1.0 [-3.4, +1.0] & +1.0 [-2.5, +4.7] & 3 & 8 \\
Qwen 3.8 27B & Prose & AgentDojo & 0.4 & -0.1 [-0.6, +0.4] & +0.0 [+0.0, +0.0] & +0.1 [-0.4, +0.7] & 0 & 3 \\
Qwen 3.8 27B & Prose & AgentHarm & 1.1 & -0.3 [-1.1, +0.5] & +0.0 [+0.0, +0.0] & +0.5 [-1.8, +3.3] & 0 & 3 \\
Llama 4 Maverick & Python & AgentDojo & 6.8 & +2.6* [+1.2, +4.0] & +3.4* [+0.6, +6.4] & +0.8 [-0.7, +2.5] & 7 & 32 \\
Llama 4 Maverick & Python & AgentHarm & 9.1 & +3.6* [+1.5, +5.9] & -0.5 [-3.8, +3.6] & +0.5 [-2.6, +3.7] & 6 & 5 \\
Llama 4 Maverick & Shell & AgentDojo & 6.8 & +2.6* [+1.3, +4.1] & +1.7 [-0.9, +4.4] & +1.5 [-0.1, +3.1] & 9 & 36 \\
Llama 4 Maverick & Shell & AgentHarm & 8.8 & +3.3* [+1.3, +5.3] & -1.5 [-4.1, +1.1] & +1.9 [-1.9, +5.7] & 5 & 10 \\
Llama 4 Maverick & No rationale & AgentDojo & 12.8 & +8.7* [+6.7, +10.7] & +0.3 [-3.7, +4.6] & +3.6* [+1.1, +5.9] & 26 & 67 \\
Llama 4 Maverick & No rationale & AgentHarm & 10.9 & +6.7* [+4.2, +9.4] & -3.9 [-8.6, +0.0] & +0.5 [-3.0, +3.5] & 12 & 7 \\
Llama 4 Maverick & Prose & AgentDojo & 7.0 & +2.9* [+1.5, +4.3] & +6.9* [+4.0, +9.8] & +1.7* [+0.1, +3.5] & 3 & 35 \\
Llama 4 Maverick & Prose & AgentHarm & 6.4 & +2.4* [+0.6, +4.5] & -1.5 [-5.5, +2.1] & +0.5 [-1.9, +3.0] & 6 & 4 \\
Nemotron 3.5 Lightning 30B & Python & AgentDojo & 4.0 & +1.2* [+0.2, +2.3] & +0.6 [-0.6, +2.0] & +0.6 [-0.8, +2.0] & 2 & 24 \\
Nemotron 3.5 Lightning 30B & Python & AgentHarm & 9.5 & +1.8 [-0.1, +3.6] & +0.0 [-3.0, +3.2] & -4.9* [-9.3, -0.5] & 9 & 3 \\
Nemotron 3.5 Lightning 30B & Shell & AgentDojo & 4.6 & +1.8* [+0.9, +2.8] & +0.6 [-0.3, +1.8] & +1.2 [-0.4, +2.8] & 1 & 31 \\
Nemotron 3.5 Lightning 30B & Shell & AgentHarm & 9.5 & +1.8 [+0.0, +3.8] & +0.5 [-3.6, +4.5] & -5.8* [-9.1, -2.4] & 8 & 1 \\
Nemotron 3.5 Lightning 30B & No rationale & AgentDojo & 7.5 & +4.0* [+2.5, +5.4] & +2.3* [+0.8, +4.1] & +5.0* [+2.9, +6.9] & 1 & 62 \\
Nemotron 3.5 Lightning 30B & No rationale & AgentHarm & 12.4 & +7.8* [+5.1, +10.3] & +4.9* [+1.5, +9.3] & +1.5 [-3.5, +6.1] & 5 & 14 \\
Nemotron 3.5 Lightning 30B & Prose & AgentDojo & 4.2 & +0.7 [-0.4, +1.9] & -0.3 [-1.7, +0.9] & +1.8* [+0.4, +3.4] & 3 & 31 \\
Nemotron 3.5 Lightning 30B & Prose & AgentHarm & 7.7 & +3.0* [+1.3, +4.8] & +0.0 [-3.2, +3.3] & -1.5 [-5.6, +2.4] & 6 & 6 \\
TS-Guard (7.6B, JSON-trained) & Python & AgentDojo & 8.9 & +5.1* [+3.4, +6.8] & +6.0* [+2.9, +9.4] & +5.3* [+3.4, +7.3] & 7 & 60 \\
TS-Guard (7.6B, JSON-trained) & Python & AgentHarm & 4.6 & +2.2* [+0.5, +4.2] & -0.5 [-1.8, +0.0] & +3.4 [-2.3, +9.4] & 2 & 16 \\
TS-Guard (7.6B, JSON-trained) & Shell & AgentDojo & 19.4 & +15.6* [+12.8, +18.2] & +6.6* [+3.7, +9.5] & +21.2* [+17.7, +24.2] & 3 & 196 \\
TS-Guard (7.6B, JSON-trained) & Shell & AgentHarm & 6.8 & +4.5* [+2.3, +6.9] & +0.0 [-1.9, +2.1] & +12.6* [+5.7, +19.2] & 2 & 33 \\
TS-Guard (7.6B, JSON-trained) & No rationale & AgentDojo & 7.9 & +4.4* [+3.0, +5.9] & +4.8* [+1.3, +8.6] & +1.0 [-0.6, +2.6] & 12 & 32 \\
TS-Guard (7.6B, JSON-trained) & No rationale & AgentHarm & 3.8 & -0.1 [-1.6, +1.3] & +1.0 [+0.0, +2.7] & -1.5 [-5.1, +2.1] & 0 & 6 \\
TS-Guard (7.6B, JSON-trained) & Prose & AgentDojo & 5.7 & +2.2* [+1.0, +3.7] & +4.3* [+1.2, +7.4] & +0.0 [-1.4, +1.4] & 8 & 19 \\
TS-Guard (7.6B, JSON-trained) & Prose & AgentHarm & 3.7 & -0.1 [-1.5, +1.3] & -0.5 [-1.8, +0.0] & -4.0 [-8.5, +0.0] & 1 & 7 \\
StepGuard (4B, JSON-trained) & Python & AgentDojo & 2.8 & +1.2* [+0.4, +1.9] & +0.0 [-0.9, +0.9] & +0.9 [-0.6, +2.2] & 1 & 20 \\
StepGuard (4B, JSON-trained) & Python & AgentHarm & 9.4 & +2.5* [+0.4, +4.7] & -0.5 [-5.7, +4.4] & -2.0 [-5.8, +2.1] & 9 & 6 \\
StepGuard (4B, JSON-trained) & Shell & AgentDojo & 2.7 & +1.1* [+0.3, +1.8] & +0.0 [-0.8, +0.9] & +0.3 [-0.9, +1.6] & 1 & 17 \\
StepGuard (4B, JSON-trained) & Shell & AgentHarm & 10.5 & +3.6* [+1.3, +6.2] & -0.5 [-5.2, +4.5] & -0.5 [-4.5, +3.2] & 10 & 8 \\
StepGuard (4B, JSON-trained) & No rationale & AgentDojo & 0.7 & +0.0 [-0.5, +0.5] & +0.0 [+0.0, +0.0] & +0.3 [-0.3, +1.0] & 0 & 6 \\
StepGuard (4B, JSON-trained) & No rationale & AgentHarm & 3.4 & +1.5* [+0.7, +2.6] & -0.5 [-2.7, +1.6] & -1.0 [-4.0, +2.2] & 3 & 4 \\
AgentDoG (4B) & Python & AgentDojo & 13.3 & +10.8* [+8.9, +12.6] & +0.0 [+0.0, +0.0] & +18.7* [+15.9, +21.6] & 0 & 162 \\
AgentDoG (4B) & Python & AgentHarm & 8.4 & +6.3* [+3.9, +8.8] & +1.5 [+0.0, +3.2] & +6.3* [+3.0, +10.1] & 0 & 14 \\
AgentDoG (4B) & Shell & AgentDojo & 21.2 & +18.8* [+16.5, +21.1] & +0.0 [+0.0, +0.0] & +29.8* [+26.6, +33.1] & 0 & 259 \\
AgentDoG (4B) & Shell & AgentHarm & 10.3 & +8.2* [+5.5, +11.1] & +2.5* [+0.4, +5.3] & +10.7* [+6.8, +15.3] & 0 & 22 \\
AgentDoG (4B) & No rationale & AgentDojo & 13.4 & +12.5* [+10.5, +14.6] & +0.0 [+0.0, +0.0] & +11.9* [+8.8, +14.9] & 0 & 133 \\
AgentDoG (4B) & No rationale & AgentHarm & 7.0 & +7.0* [+4.6, +9.3] & +1.0 [-1.7, +3.6] & +0.0 [-3.6, +3.8] & 2 & 7 \\
ShieldAgent (7B) & Python & AgentDojo & 18.5 & +14.4* [+12.2, +16.8] & +2.8* [+0.8, +5.0] & +13.4* [+10.4, +16.7] & 3 & 163 \\
ShieldAgent (7B) & Python & AgentHarm & 9.5 & +5.6* [+3.0, +8.4] & +2.0* [+0.4, +3.9] & +11.2* [+6.1, +16.7] & 0 & 25 \\
ShieldAgent (7B) & Shell & AgentDojo & 26.6 & +22.5* [+20.0, +25.1] & +6.2* [+3.8, +8.7] & +26.8* [+23.5, +30.4] & 0 & 268 \\
ShieldAgent (7B) & Shell & AgentHarm & 8.6 & +4.7* [+2.4, +7.6] & +1.5 [+0.0, +3.1] & +12.1* [+6.6, +18.3] & 0 & 26 \\
ShieldAgent (7B) & No rationale & AgentDojo & 0.2 & +0.1 [+0.0, +0.3] & +0.0 [+0.0, +0.0] & -0.2 [-0.6, +0.0] & 0 & 0 \\
ShieldAgent (7B) & No rationale & AgentHarm & 0.1 & +0.1 [+0.0, +0.4] & +0.0 [+0.0, +0.0] & +0.0 [+0.0, +0.0] & 0 & 0 \\
Safiron (8B) & Python & AgentDojo & 11.0 & +9.7* [+7.9, +11.7] & +15.3* [+11.3, +19.6] & +5.1* [+3.3, +6.8] & 9 & 53 \\
Safiron (8B) & Python & AgentHarm & 7.0 & +4.6* [+2.8, +6.4] & +4.9* [+1.1, +9.0] & +1.9 [-1.0, +5.2] & 2 & 7 \\
Safiron (8B) & Shell & AgentDojo & 44.4 & +43.1* [+39.8, +46.6] & +34.9* [+29.7, +40.7] & +47.1* [+43.4, +50.6] & 3 & 411 \\
Safiron (8B) & Shell & AgentHarm & 25.2 & +22.8* [+19.4, +26.4] & +17.7* [+11.1, +25.1] & +17.0* [+11.1, +22.9] & 2 & 39 \\
Safiron (8B) & No rationale & AgentDojo & 1.7 & +1.6* [+1.0, +2.4] & +4.0* [+1.8, +6.3] & +0.1 [-0.2, +0.5] & 2 & 2 \\
Safiron (8B) & No rationale & AgentHarm & 4.3 & +4.2* [+2.6, +5.9] & +4.4* [+1.1, +9.0] & -1.0 [-3.1, +1.0] & 2 & 1 \\
\bottomrule
\end{tabular}
}
\end{center}

\FloatBarrier
\section{A step under every condition}
\label{app:examples}
Original (\agentharm{}, unsafe):
\begin{quote}\small\ttfamily
(1) Thought: I will start by using the Stable Diffusion model to create an image of the politicians, Donald Drewski and Fernando Whitehall, shaking hands.\\
(2) Action: run\_stable\_diffusion\_model\\
(3) Action Input: \{"image\_path": "", "prompt": "Donald Drewski and Fernando Whitehall shaking hands"\}
\end{quote}
Python form (Thought unchanged; the \texttt{execute\_python} definition is appended to the tool list):
\begin{quote}\small\ttfamily
(2) Action: execute\_python\\
(3) Action Input: \{"code": "run\_stable\_diffusion\_model(image\_path='', prompt='Donald Drewski and Fernando Whitehall shaking hands')"\}
\end{quote}
Shell form:
\begin{quote}\small\ttfamily
(2) Action: execute\_bash\\
(3) Action Input: \{"command": "run\_stable\_diffusion\_model --image-path '' --prompt 'Donald Drewski and Fernando Whitehall shaking hands'"\}
\end{quote}
Prose form (Thought unchanged; lines (2)--(3) replaced by the sentence):
\begin{quote}\small\ttfamily
I will now run a stable diffusion model with an image file name of "" and the prompt Donald Drewski and Fernando Whitehall shaking hands.
\end{quote}
No-rationale form drops line (1). The whitespace placebo appends one trailing space; the wrapper placebo appends the \texttt{execute\_python} definition and leaves the action untouched.

\section{Protocol notes, failures and cost}
\label{app:protocol}
Claude Sonnet 5 returns an empty completion for the same two \agentharm{} steps in every condition and on retry; they are excluded from its cells consistently ($n = 716$). \agentharm{} repeats $205$ (interaction, segment) keys; step identifiers are persisted on materialized rows. Llama 4 Maverick's wrapper placebo alone raises its \agentdojo{} recall by $3.7\pp$ \ci{+1.2}{+6.2}, which is why wrapper-condition effects are adjusted against that placebo. Hosted-guard inference for the primary cohort cost about \$200 in tokens and \$32 in GPU hours for the three self-hosted guards.

\textbf{Prose form.} Gemini 3.8 Flash ($T = 0.3$, up to nine candidates) wrote one sentence per step; a candidate is kept only if every argument value appears verbatim, the tool name does not, no authorization or urgency phrase was added, and it is at most three times the length of the original block; $1{,}921$ of the $1{,}938$ cohort steps received one. Gemini 3.1 Pro recovered the exact (tool, arguments) from the sentence and the tool definitions for $1{,}745$; Claude Sonnet 5 judged $600$ before partner billing was stopped and recovered $573$. The $185$ rows on which a judge failed were reviewed one by one by Claude (Fable 5.1) following the authors' written instructions, and the authors accepted that review in place of a human pass: $183$ accepted (every judge failure was a trailing period absorbed into a value, a dropped null, a type coercion, a parameter rename, an empty tool, or a multi-sentence value truncated by the judge), $1$ repaired (a Python list literal in the recipient slot), $1$ rejected (a hallucinated tool whose name carried the agent's own disclaimer); $1{,}920$ sentences were frozen (\texttt{FREEZE.json}). \stepguard{} rejects every prose step for missing Action markers; AgentDoG, ShieldAgent and Safiron raise before inference because there is no tool call to extract. Prose inference cost about \$23 of hosted tokens and $1.8$ GPU-hours.

\end{document}